\documentclass[aps,prb,twocolumn,superscriptaddress,bibliography]{revtex4-2}

\usepackage{amsmath}
\usepackage{amssymb}
\usepackage{graphicx}
\usepackage{epstopdf} 
\AppendGraphicsExtensions{.tif}

\usepackage{xcolor}
\usepackage{comment}
\usepackage{soul}

\makeatletter
\newcommand*{\addFileDependency}[1]{
  \typeout{(#1)}
  \@addtofilelist{#1}
  \IfFileExists{#1}{}{\typeout{No file #1.}}
}
\makeatother

\newcommand{\NCTO} {Na$_2$Co$_2$TeO$_6$}

\begin{document}


\title{Out-of-plane magnetic phase diagram of Kitaev quantum spin liquid candidate Na$_2$Co$_2$TeO$_6$}



\author{Shengzhi Zhang}
\affiliation{National High Magnetic Field Laboratory, Los Alamos National Laboratory, Los Alamos, New Mexico 87545, USA.}

\author{Sangyun Lee}
\affiliation{National High Magnetic Field Laboratory, Los Alamos National Laboratory, Los Alamos, New Mexico 87545, USA.}
\author{Eric Brosha}
\affiliation{Los Alamos National Laboratory, Los Alamos, New Mexico 87545, USA.}
\author{Qing Huang}
\affiliation{Department of Physics, University of Tennessee, Knoxville, Tennessee 37996, USA.}
\author{Haidong Zhou}
\affiliation{Department of Physics, University of Tennessee, Knoxville, Tennessee 37996, USA.}
\author{Vivien S. Zapf}
\affiliation{National High Magnetic Field Laboratory, Los Alamos National Laboratory, Los Alamos, New Mexico 87545, USA.}
\author{Minseong Lee}
\email{ml10k@lanl.gov}
\affiliation{National High Magnetic Field Laboratory, Los Alamos National Laboratory, Los Alamos, New Mexico 87545, USA.}

\date{\today}


\date{\today}

\begin{abstract}
We have investigated the magnetic properties and mapped out the phase diagram of the honeycomb magnet \NCTO{} with Co 3$d^{7}$ in out-of-plane magnetic fields. This material has previously been proposed to show nearest-neighbor Kitaev interactions between Co spins and maybe even Kitaev quantum spin liquid behavior in high fields. At low magnetic fields, we observe a thermal phase transition at $T_{\text{N}}$ = 27 K, transitioning from a paramagnetic state to a canonical ferrimagnetic state. Under the application of magnetic fields, a spin flop-like phase transition occurred before saturation of \textit{J} = 1/2 between 10 K and $T_{\text{N}}$. Below 10 K, a peak-dip-peak structure emerges between 10 and 17 T in the magnetic susceptibility ($dM/dH$) before the magnetic saturation, reminiscent of magnetic plateau behavior. The measurement of the magnetocaloric effect also shows dip-peak-dip behavior in this field range. Our data can be explained by an XXZ model with a single ion anisotropy and possibly small Kitaev and $\Gamma$ exchange interactions. We also unambiguously determined the magnetization saturation field that helps constrain the energy scale of the exchange interactions. 
\end{abstract}


\maketitle

\section{Introduction}

The honeycomb lattice with bond-dependent nearest-neighbor exchange interactions was proposed by Kitaev as a rare example of frustration leading to quantum spin liquid behavior with an analytical solution.\cite{Kitaev2006anyons}
By preserving time reversal symmetry (TRS) and ensuring comparable magnitudes of the three nearest neighbor exchange interactions, the ground state exhibits non-zero spin-spin correlation exclusively between nearest neighbors' spins and finite multispin correlations —a characteristic indicative of a Kitaev quantum spin liquid (KQSL) ground state. Such a ground state is expected to be gapless.
Moreover, it was demonstrated that appropriate perturbations, such as a magnetic field {\it perpendicular} to the honeycomb lattice, can open an energy gap, transforming the ground state into a non-Abelian topological phase. This topological phase holds unprecedented potential for realizing fault-tolerant quantum computations \cite{nayak2008non,kitaev2003fault}.


Jackeli and Khaliullin later demonstrated \cite{jackeli2009mott} that materials can exhibit Kitaev's proposed bond-dependent Ising interactions through spin-orbit-entangled Kramers pairs, facilitated by strong spin-orbit coupling ($\lambda$) and crystal electric field effects. Subsequent research revealed the presence of symmetric off-diagonal exchange $\Gamma$ and additional interactions, such as $\Gamma'$ and a single ion anisotropy term, $A_{c}$ , due to deviations from the perfect 90$^{\circ}$ bond angle and direct overlap of $d$ orbitals \cite{chaloupka2010kitaev,rau2014generic, katukuri2014kitaev,stavropoulos2021magnetic}.
These theoretical studies have identified numerous materials possessing low spin (LS) $d^{5}$ transition metal ions with strong $\lambda$, resulting in a $J_{\text{eff}} = 1/2$ Kramers doublet. Examples include $A_{2}$IrO$_{3}$ ($A$ = Na, Li) \cite{winter2017models,takagi2019concept, chaloupka2010kitaev} and $\alpha$-RuCl$_{3}$ \cite{plumb2014alpha, do2017majorana}, where the latter has garnered significant attention. Although $\alpha$-RuCl$_{3}$ orders below 7 K due to various additional magnetic exchange terms, it has been demonstrated that an in-plane magnetic field destabilizes the zigzag spin structure in the ground state \cite{zheng2017gapless, majumder2015anisotropic, kubota2015successive, leahy2017anomalous, baek2017evidence, sears2017phase, wolter2017field}. This disruption induces a new phase reminiscent of a chiral spin liquid, which may be supported by a remarkable half-integer quantized thermal Hall effect \cite{kasahara2022quantized}.

In the quest for materials where Kitaev interactions could dominate, 
recent calculations have considered the 3$d^7$ high spin (HS) state.\cite{liu2018pseudospin, liu2021towards, sano2018kitaev, liu2020kitaev, motome2020materials}. 
Introducing two extra electrons in $e_{g}$ orbitals creates Heisenberg exchange interactions opposing those from $t_{2g}$ orbitals, thus suppressing the overall Heisenberg exchange interaction. The Co$^{2+}$ ion with a $d^7$ configuration in an octahedral crystal field fulfills the conditions for the extended $d^7$ HS mechanism. As a result, cobalt-based honeycomb lattices like \NCTO{}, Na$_3$Co$_2$SbO$_6$ \cite{viciu2007structure}, and BaCo$_2$(AsO$_4$)$_2$ \cite{regnault1977magnetic, zhong2020weak} were then considered as potential Kitaev materials. However, this proposition faces challenges as achieving another necessary condition—strong metal-ligand orbital hybridization—is difficult in $3d$ compounds due to a large charge transfer energy gap \cite{das2021xy, maksimov2022ab, winter2022magnetic}.

On the other hand, cobaltites, having a unique global Z axis, are traditionally described by the bond-independent XXZ model. The first-order effect of $\lambda$ stabilizes the $j = 1/2$ state as the ground state in this framework. Trigonal distortion induces a distinct spin magnitude along the Z axis compared to the X and Y directions. When the trigonal elongation surpasses half of $\lambda$, the second-order $\lambda$ effect between $j = 1/2$ and $j = 3/2$ states leads to an easy-planar type single-ion anisotropy \cite{lines1963magnetic, oguchi1965theory}.

In \NCTO{} 
\cite{viciu2007structure, bera2017zigzag, lin2022evidence, zhang2023electronic, xiang2023disorder, sanders2022dominant}, the $3d^7$ Co$^{2+}$ ions form a honeycomb lattice in the $ab$-plane with nearly 90$^{\circ}$ Co$^{2+}$-O$^{2-}$-Co$^{2+}$ bond angles and comparable $\lambda$ to other energy scales \cite{kim2021antiferromagnetic}. These characteristics position \NCTO{} as an  candidate for a KQSL based on conditions suppressing Heisenberg and symmetric off-diagonal terms \cite{liu2021towards}. However, several recent works find that a third nearest-neighbor Heisenberg interaction across the honeycomb can dominate, thereby probably breaking the condition for a KQSL \cite{Kruger23,Winter22,Yao22}. For in-plane fields, a THz spectroscopy study finds a continuum of excitations\cite{pilch2023field} that could be consistent with a spin liquid, while another study finds that disordered Na vacancies could produce this continuum \cite{xiang2023disorder}. Several thermal conductivity studies have investigated the question of whether a KQSL could be consistent with the data for in-plane magnetic fields. \cite{Guang23,Guang24,hong2021strongly}

In this study, we explore the magnetic properties of \NCTO{} with a magnetic field applied along the $c$-axis, perpendicular to the honeycomb plane. We construct a magnetic phase diagram through various experimental probes, including dc magnetic susceptibility, magnetization, and magnetocaloric effect measurements. These measurement allow us to unambiguously determine the saturation magnetic field around 43 T that reveals pronounced anisotropy between in-plane and out-of-plane magnetic properties of \NCTO{}. Interestingly, we may have observed a magnetic plateau behavior at intermediate magnetic fields between 10 and 17 T only below 10 K, indicating the presence of magnetic frustration. This observation hints at the possible existence of Kitaev exchange interactions in the \NCTO{} compound.

\section{Experiments}
The single crystal synthesis method is the same as presented in our previous work \cite{zhang2023electronic}, forming in a \textit{P}6$_{3}$22 (No. 182) space group. The honeycomb layer is formed by edge-sharing CoO$_{6}$ octahedra ($S$ = 3/2, $L_{\text{eff}}$ = 1/2), with each layer separated by nonmagnetic Na atoms. Consistent magnetic susceptibilities and specific heat measurements across various single crystalline samples employed in this study validate the uniform crystal quality and consistent magnetic/thermodynamic properties across all samples.

We conducted temperature-dependent dc magnetic susceptibility measurements using the vibrating sample magnetometry (VSM) method in a 14 T Quantum Design Physical Property Measurement System (PPMS), with the magnetic field aligned along the $c$-axis. Additionally, field-dependent magnetization experiments were carried out in a 65 T millisecond short-pulse magnet, reaching up to 60 T along the $c$-axis.\cite{zhang2023electronic} The raw voltage data from the magnetic induction coil were subsequently calibrated using the PPMS VSM data acquired from the same crystals and then converted to magnetization.

The magnetocaloric effect, measured as the sample temperature change versus the magnetic field, was conducted in pulsed magnetic fields. To establish a robust thermal connection between the sample and the thermometer on millisecond timescales in pulsed fields, a semiconducting AuGe thin film was directly deposited onto the sample as a thermometer. The film was deposited through RF magnetron sputtering at 40 mTorr pressure of ultra-high purity Ar gas for 60 minutes with 100 W power. Subsequently, Au contact pads were deposited on top of the AuGe film using a shadow mask, leaving a stripe of AuGe uncovered. For measuring the thermometer resistance in pulsed fields, a custom digital lock-in method with a 100 kHz source current was employed, utilizing the four-point method as typically done at the National High Magnetic Field Laboratory Pulsed Field Facility. A detailed schematic of the setup is available in the supplemental materials of Ref. \cite{zhang2023electronic}. The thermometer underwent calibration in thermalized conditions with helium exchange gas to establish the resistance versus temperature relationship, and an identical reference thermometer (AuGe on glass) was employed for the magnetoresistance calibration.



\section{Results}

    
\subsection{dc magnetic susceptibility vs. temperature}
\begin{figure}
\includegraphics[width=\linewidth]{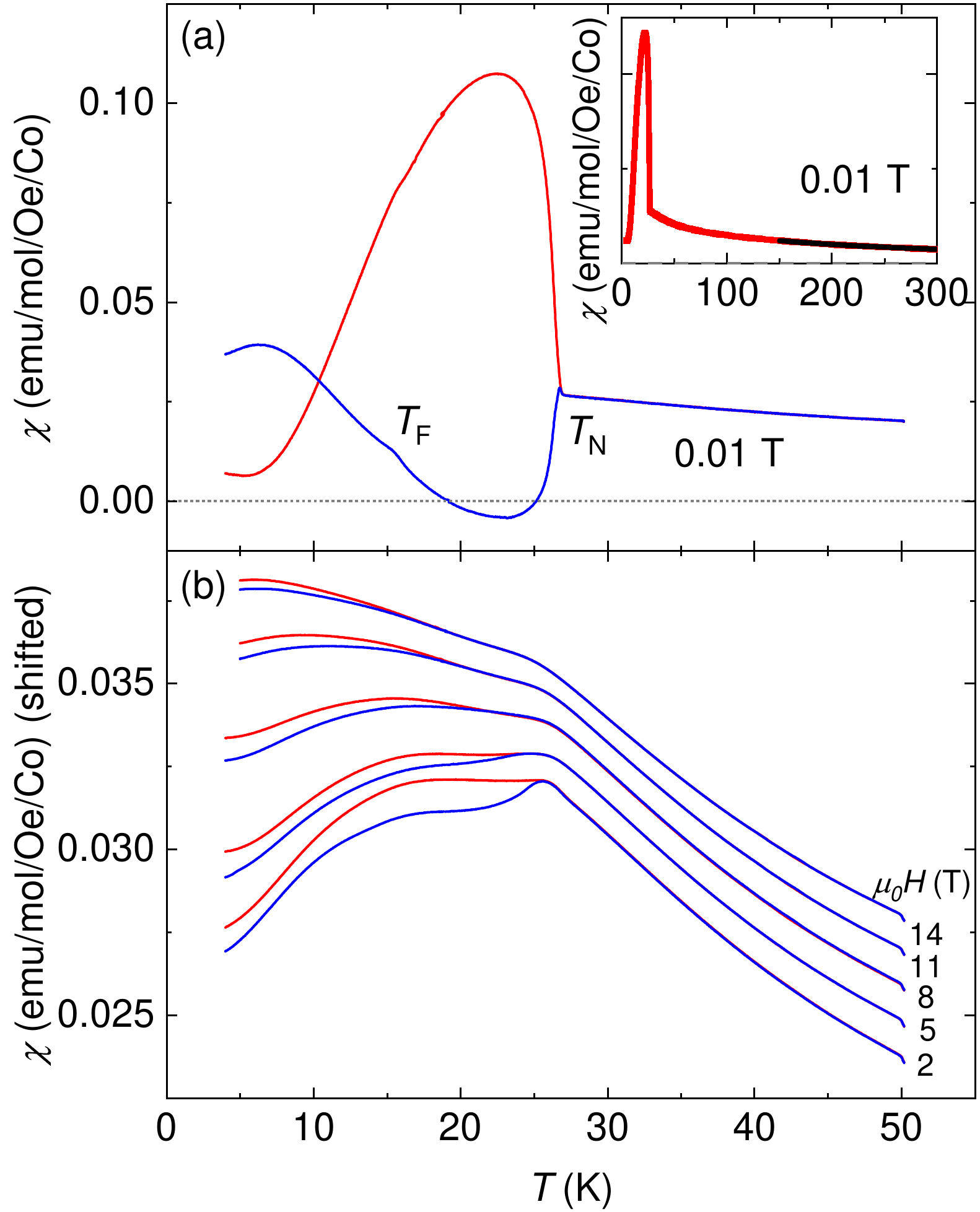}%
\caption{Red and blue curves represent field cooled and zero-field cooled magnetic susceptibility, respectively, as a function of temperature measured at (a) 0.01 T along $c$-axis and (b) higher magnetic fields with a constant shift 0.00125 emu/mol/Oe/Co applied for clarity. The inset of (a) shows $\chi$ vs. \textit{T} up to 300 K at 0.01 T. The thick red curve is the experimental data and the thin black curve is the Curie-Weiss fit. \label{MvsT}}
\end{figure}

The zero-field cooled (ZFC) and field cooled (FC) dc magnetic susceptibility as a function of temperature ($\chi_{dc}^{ZFC}(T)$ and $\chi_{dc}^{FC}(T)$) at various out-of-plane magnetic fields from 0.01 T to 14 T are shown in Fig.~\ref{MvsT} (a) and (b). In all magnetic fields, a clear anomaly and a strong bifurcation between ZFC and FC at around $T_{\text{N}}$ is observed. At $H = 0.01$ T (Fig.~\ref{MvsT} (a)), the $\chi_{dc}^{ZFC}(T)$ magnetic susceptibility becomes negative between 19 and 26 K while $\chi_{dc}^{FC}(T)$ strongly diverges at the same temperature range. This suggests that the Co$^{2+}$ moments align ferrimagnetically along $c$-axis below $T_{\text{N}}$, consistent with the previous observation in Ref.~\cite{yao2020ferrimagnetism}. Another critical temperature is observed as a weak peak and labeled as $T_{\text{F}}$ in Fig.~\ref{MvsT} (a). Both features broaden out with increasing magnetic field and $T_{\text{F}}$ quickly become indiscernible above 1.2 T. $T_{\text{N}}$, on the other hand, decreases slowly from 0 to 14 T, as shown in Fig.~\ref{MvsT} (b). Below 5 K, the discrepancy between $\chi_{dc}^{ZFC}(T)$ and $\chi_{dc}^{FC}(T)$ begins close to $T_{\text{N}}$. However, the temperature at which ZFC and FC begin to diverge is suppressed toward lower temperatures. Therefore, the ferrimagnetic domains are unpinned when thermal fluctuations supply sufficient energy to allow domain walls to move. Na-occupation disorder that strongly affects the spin dynamics could be a source of the strong domain pinning \cite{xiang2023disorder}.

The inset of Fig.~\ref{MvsT} (a) shows $\chi_{dc}^{FC}(T)$ measured at 0.01 T as a representative curve for Curie-Weiss fit. All magnetic susceptibilities measured up to 14 T match at 30 K or higher. We fit the curve between 150 K and 300 K with $\chi = \frac{C}{T-\Theta_{\text{CW}}} + \chi_{0}$. We have extracted $\Theta_{\text{CW}} = -113$ K, $\mu_{\text{eff}} = 5.13 \mu_{\text{B}}$, and $\chi_0 = -0.00052$ emu/mol/Oe/Co. The small $\chi_{0}$ takes care of the various background from the sample holder and glue used in the experiment. 
On the other hand, the in-plane parameters are extracted in a similar manner to be $\mu_{eff} = 5.7\mu_B$, $\Theta_{\text{CW}} = 2.11$ K whose sign is opposite from that of out-of-plane, and $\chi_0 = -0.002$ emu/mol/Oe/Co (see appendix). This anisotropic Curie-Weiss behavior can be a result of anisotropic magnetic exchange such as Kitaev interactions\cite{Li21}, or Heisenberg magnetic exchange plus easy-plane single-ion anisotropy \cite{Fernengel79,Zapf16}, or both.

\subsection{High field magnetization}
\begin{figure}
\includegraphics[width=\linewidth]{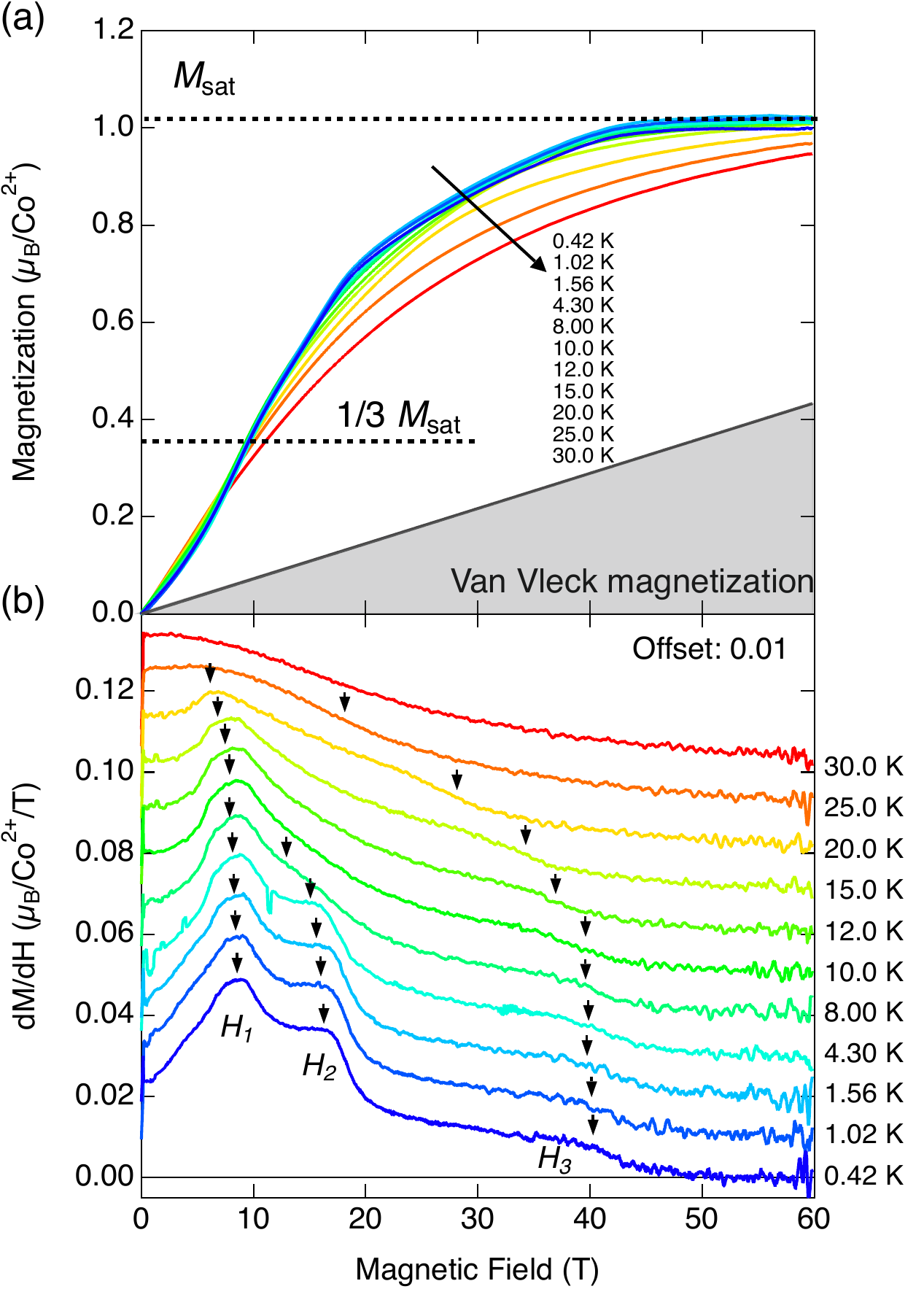}%
\caption{(a) Magnetization of \NCTO{} measured at various temperatures in pulsed field after the subtraction of Van Vleck magnetization. The subtracted Van Vleck magnetization is shown as grey shadow area. The raw data without Van Vleck magnetization subtraction can be found in S. I. (b) The derivative of the magnetization shown in (a) with magnetic field. The offset of 0.01 $\mu_{B}/\text{Co}^{2+}/\text{T}$ were applied for clarity. Black arrows indicate the phase transition fields.}
\label{MvsH}
\end{figure}

Fig.~\ref{MvsH} illustrates the pulsed field dependence of magnetization, $M(H)$, and $dM/dH$ at various temperatures. Note that in this figure we have already removed a Van Vleck effect contribution to the magnetization. While $M(H)$ appears to increase smoothly with the magnetic field, $dM/dH$ reveals a rich structure in the magnetization. Above $T_{\text{N}}$, $M(H)$ follows the Brillouin function, consistent with the paramagnetic phase. When the temperature is between $T_{\text{N}}$ and 10 K, a single strong peak in $dM/dH$ emerges around 9 T, resembling the behavior of spin-flop transitions. We mark this first phase transition field as $H_{1}$ The concave behavior in $M(H)$ below $H_{1}$ suggests that the magnetism is effectively two-dimensional \cite{goddard2008experimentally}, indicating negligible interlayer coupling compared to intralayer coupling.

As we further lower the temperature, a peak-dip-peak structure in $dM/dH$ between $H_{\text{1}}$ and 17 T ($H_{\text{2}}$) emerges, reminiscent of a plateau behavior. The field of the anomaly at $H_{\text{1}}$ is largely temperature-independent but gradually disappears as the temperature approaches $T_{\text{N}}$. However, the anomaly at $H_{\text{2}}$ is robust against magnetic field at very low temperatures but is quickly suppressed above 10 K. Lastly, although weaker than the first two anomalies, another anomaly is observed around 40 T, denoted as $H_{\text{3}}$. As we will discuss later, these anomalies are smoothly connected to the $T_{\text{N}}$ phase boundary; therefore, we assign $H_{\text{3}}$ as the saturation field. The additional field-induced magnetization after $H_{\text{3}}$ is attributed to the Van Vleck paramagnetic contribution with a slope of $\sim 0.01 , \mu_{B}/\text{T}/\text{Co}$, comparable to other cobaltites \cite{Lee2014VV, susuki2013magnetization}. From the dotted line, we obtain a saturation magnetization of $\sim$ 1.01 $\mu_B$/Co. Substituting this value into $M = g\mu_B J$, where $J = 1/2$, we estimate the $g$-factor to be $\sim$ 2.02, consistent with the value obtained from electron spin resonance experiments \cite{lin2021field}, and the Zeeman energy for the saturation is about $2.3$ meV/Co. This saturation magnetization contrasts with the $\mu_{eff}$ = 5.13 $\mu_B$ obtained from the Curie-Weiss fit above. The discrepancy is due to the fact that the Curie Weiss fit, obtained between 150 and 300 K, is seeing the excited \textit{J} = 3/2 state.

\subsection{Magnetocaloric effect}

\begin{figure}
\includegraphics[width=\linewidth]{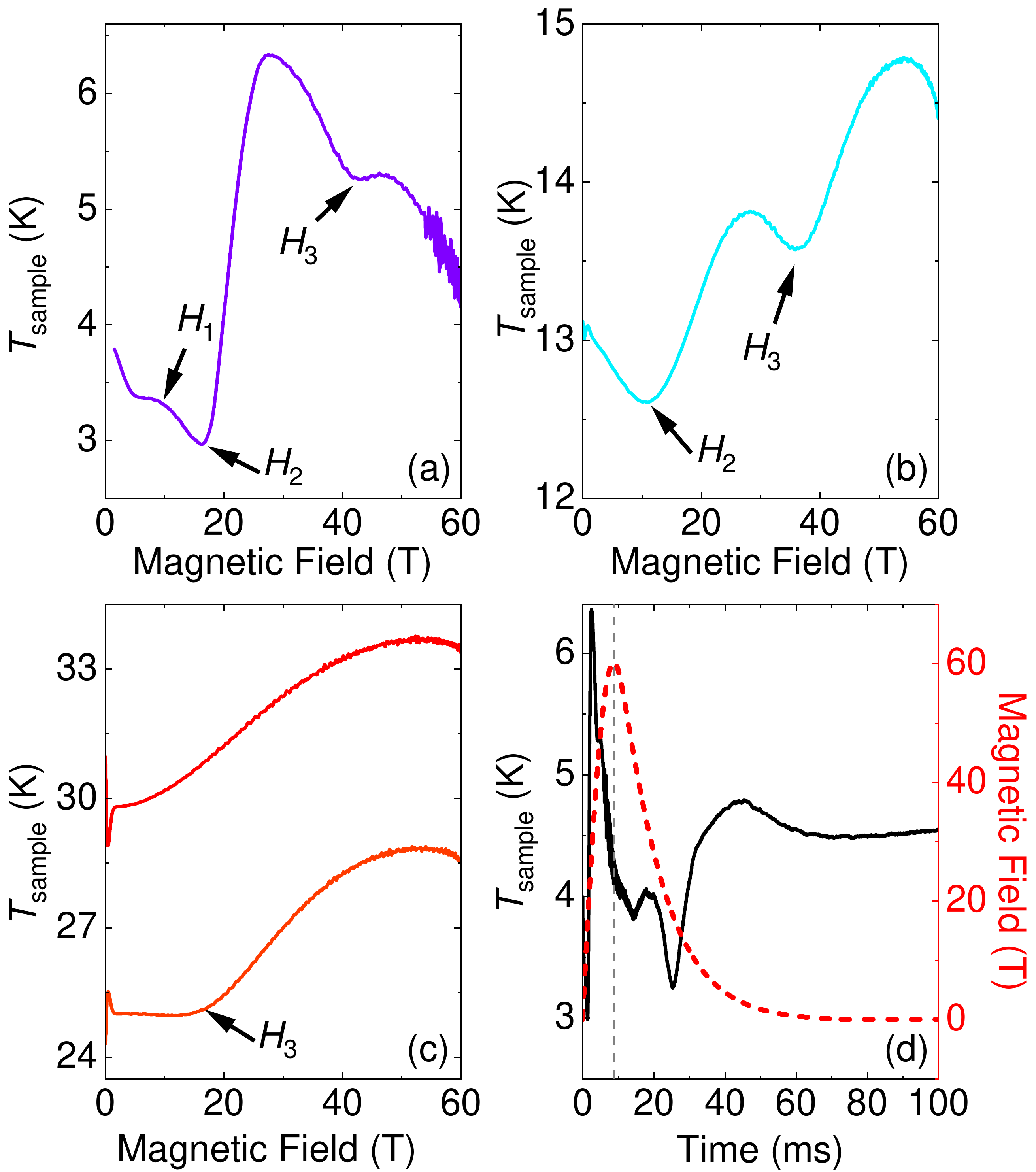}%
\caption{(a)-(c) Selected sample temperature as a function of magnetic field (up-sweep) data. Arrows indicate the critical fields observed in each curve at different temperatures. (d) The same data curve as (a) but as a function of time and with the field-down sweep. The red dotted line shows the magnetic field profile. The grey dashed line indicates where maximum magnetic field is. \label{MC}}
\end{figure}

The magnetocaloric effect, i.e. temperature of \NCTO{} measured in millisecond pulsed fields vs magnetic field are depicted in Fig.~\ref{MC} (a)-(c). The temperature measured above $T_{\text{N}}$ (top curve in Fig.~\ref{MC} (c)) gradually increases initially from 0 T to 50 T and starts decreasing at higher fields. The temperature of a paramagnetic sample is expected to continuously increase with the magnetic field, aligning spins under \textit{adiabatic} conditions. Hence, our data suggest that the relaxation time of the sample in this environment is less than 9 ms (the time-frame within which these data were taken). The relaxation time, proportional to the heat capacity of the sample, decreases at lower temperatures, resulting in a shorter relaxation time. Consequently, non-adiabatic conditions are anticipated for the curves measured below $T_{\text{N}}$. As illustrated in Fig.~\ref{MC} (d), the exponential decay of the sample temperature across the dotted grey line denoting the maximum magnetic field, and the discrepancy between the field sweep up and down curves, support the notion that the measurement conditions lie between equilibrium and adiabatic conditions. However, we could clearly identify anomalies of peaks and dips denoted with black arrows in Fig.~\ref{MC} (a) - (c), corresponding to the anomalies observed in the magnetization data. The first two low-field features at $H_{1}$ and $H_{2}$ are likely captured under adiabatic conditions due to the ultra-fast field sweep, while the feature at $H_{3}$ might contain a significant cooling background due to interaction with the environment at a lower field sweep rate. Additional explanation of how the sample is in adiabatic condition at lower fields and semi-adiabatic condition around maximum field is given in the Appendix.

Observing how the temperature changes in the sample under adiabatic conditions provides insights into the nature of the phase transition. Below $H_{1}$, the temperature slightly increases, forming a peak around 10 T, then cools back again before rising sharply around $H_{2}$, showing a dip-peak-dip structure. Assuming that near $H_{1}$ and $H_{2}$ the close-to-adiabatic condition is fulfilled, this behavior aligns with the magnetocaloric effect near a magnetic plateau \cite{morita2022isothermal, antonio2023magnetocaloric}. At around $H_{3}$, the temperature briefly increases and gradually drops afterward. Since $H_{3}$ is the saturation field, we expect the temperature to increase as the magnetic gap widens. Therefore, we believe that the $H_{3}$ feature is a combination of the gradually increasing temperature due to the magnetic entropy change with the strong cooling effect due to interaction with the environment. As shown in Fig.~\ref{PD}, $H_{1}$ changes little with increasing temperature; on the other hand, $H_{2}$ is quickly suppressed with increasing temperature and merges with $H_{1}$ above 10 K. $H_{3}$ gradually decreases with increasing temperature and smoothly connects to $T_{\text{N}}$ at zero magnetic field, indicating that $H_{3}$ is the saturation magnetic field for the \textit{J} = 1/2 state.



\section{Discussion}

\begin{figure}
\includegraphics[width=\linewidth]{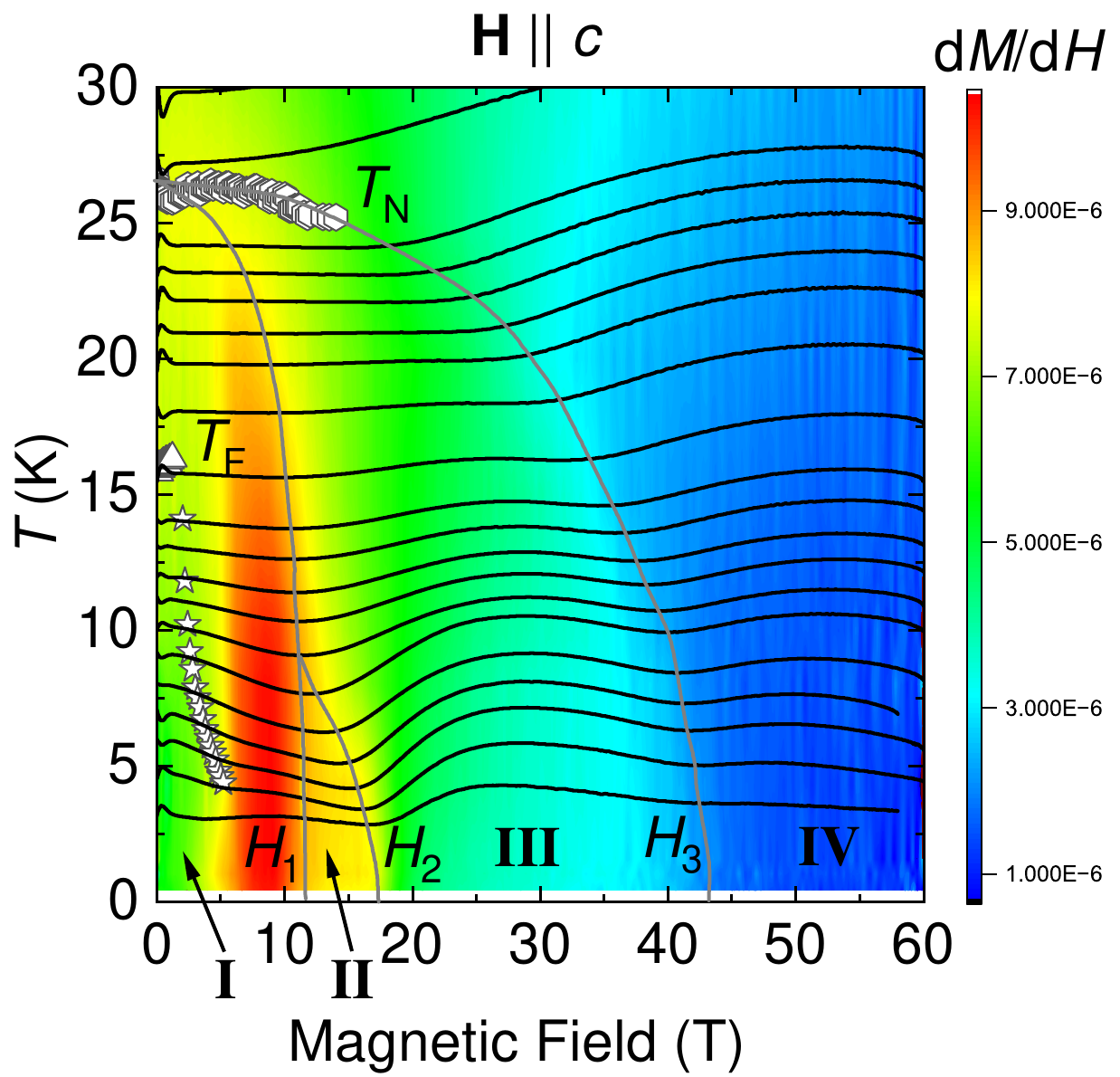}%
\caption{\textit{T-H} phase diagram of \NCTO{} with \textit{H} $\parallel$ \textit{c}-axis. Four phases are observed (I to IV). The colored contour plot is based on \textit{dM/dH} data. The open symbols are from the \textit{$\chi$} vs. \textit{T} data in which the open triangles represent $T_{\text{F}}$ and the open stars are explained in the appendix. The solid black lines are sample temperatures as a function of magnetic field from magnetocaloric effect measurements. The solid grey lines (labeled by $H_{\text{1}}$ to $H_{\text{3}}$) are guide to the eyes tracing the peaks/dips in the magnetocaloric effect data. A dashed line as part of $H_{\text{1}}$ represents an estimation for part of this phase boundary. Note that error bars are added but most of them are smaller than the data points. \label{PD}}
\end{figure}

We present a temperature-magnetic field ($T - H$) phase diagram of \NCTO{} with magnetic fields applied along the \textit{c}-axis in Fig.~\ref{PD}. From the matching features between the magnetocaloric effect data and the field-dependent magnetization data, we found four phases (I--IV) with clear phase boundaries ($H_{\text{1}}$--$H_{\text{3}}$) in addition to the paramagnetic state above $T_{\text{N}}$ and $H_{3}$. $H_{\text{1}}$ is defined by a peak in sample temperature curves coinciding with the local maximum curvature in the field-dependent magnetization. $H_{\text{2}}$ is defined at a dip in sample temperature curves matching with a shoulder in $dM/dH$. $H_{\text{2}}$ joins $H_{\text{1}}$ at around 10 K as the shoulder in $dM/dH$ disappears, and the peak-dip feature in sample temperature curves merge together. $H_{\text{1}}$ ends at $T_{\text{N}} \approx$ 27 K at 0 T, which is also observed in temperature-dependent magnetic susceptibility and specific heat measurements. $H_{\text{3}}$ is defined by another shoulder in $dM/dH$ along with a dip in magnetocaloric effect data. It connects smoothly to $T_{\text{N}}$, indicating that it is the saturation magnetic field of \textit{J} = 1/2. Another critical temperature is observed in temperature-dependent magnetic susceptibility and specific heat measurements at $T_{\text{N}} \approx$ 15 K, similar to a critical temperature observed in in-plane measurements \cite{zhang2023electronic}. The critical temperature shown in Fig.~\ref{PD} as empty stars is observed in zero-field cooled magnetic susceptibility measurements as a peak (see Appendix) but cannot be unambiguously seen in the pulsed field data since this feature is within the regime of the initial noise below 5 T from the pulsed field firing.

The dc magnetic susceptibility data show a canonical ferrimagnetic nature of phase I, consistent with previous results \cite{yao2020ferrimagnetism}. It requires a canted moment along the c-axis to explain the data. Microscopically, the $\Gamma$ term may induce out-of-plane spin components. In phase III, the magnetization smoothly increases with the magnetic field; therefore, we can imagine spins gradually canting towards the magnetic field directions in this phase. All spins are aligned along the field direction in phase IV. We successfully extracted the $g$-factor anisotropy, which is about 2. This is significantly less than the in-plane $g$-factor anisotropy, approximately 4.3, from our previous study \cite{zhang2023electronic} and another EPR study \cite{lin2021field}.

The evolution of single-ion properties, including in-plane and out-of-plane $g$-factor anisotropy, with trigonal distortion has been studied for Co$^{2+}$ under a octahedral crystal field \cite{lines1963magnetic, oguchi1965theory, winter2022magnetic}. Our $g$-factor anisotropy suggests that the trigonal distortion is approximately $0.6\lambda$, which is consistent with other experimental findings \cite{kim2021antiferromagnetic}. The small energy gap of $\sim 6$ meV between $J = 1/2$ and $J = 3/2$, given that $\lambda \sim 30$ meV \cite{kim2021antiferromagnetic}, can induce easy-planar type single-ion anisotropy. The XXZ model combined with the single-ion anisotropy shows a nearly temperature independent single phase transition \cite{holtschneider2008uniaxially} before magnetization saturation, which well describes the magnetic phase diagram between 10 K and $T_{\text{N}}$. This implies that a suitable spin model for \NCTO{} is close to the XXZ model with a residual easy-plane single-ion anisotropy \cite{winter2022magnetic}, in addition to other symmetrically allowed exchange interactions.  

Interestingly, we have observed a magnetic plateau-like behavior, occurring around 1/3 of the saturation magnetization below 10 K (Phase II). Magnetic plateaus are a phenomenon observed in many geometrically frustrated systems \cite{lee2024magnetic, shirata2012experimental, susuki2013magnetization}, mainly due to the order-by-disorder effect. Recently, a 1/3 plateau in a nickel-based honeycomb lattice with a magnetic field perpendicular to the spin ordering axis has been reported. Intriguingly, the Kitaev exchange interaction was identified as the origin of the 1/3 plateau \cite{shangguan2023one}. Their Hamiltonian also contains first nearest, third nearest, and Kitaev exchange interactions with single-ion anisotropy. Another possible scenario is that further neighbor exchange interactions give rise to a small geometrical frustration within the the XXZ model.

Thus our results could be consistent with a Kitaev interaction also being present, producing the 1/3 magnetization plateau, which is washed out as the thermal energy overcomes the energy scale of the Kitaev interactions above 10 K, as observed in Ref.~\cite{shangguan2023one}.

Therefore, a potential spin Hamiltonian for \NCTO{} consists of a dominant XXZ model, plus a weak Kitaev exchange interaction that give rise to magnetic frustration, and a $\Gamma$ term making the spins canted along the \textit{c}-axis. The Kitaev interaction is predicted to be weak because it is washed out as the thermal energy overcomes the energy scale of the Kitaev interactions above 10 K. One potential reason for the weak plateau is that the magnetic field direction is not perfectly perpendicular to the spin ordering axis due to potential spin canting.

\section{Conclusion}
We have constructed a comprehensive out-of-plane \textit{T}-\textit{H} phase diagram up to magnetic field saturation  of \textit{J} = 1/2 of a Kitaev quantum spin liquid candidate \NCTO{} via a combination of dc magnetization and magnetocaloric effect measurements. An additional 0.4 $\mu_B$ of magnetization appears by 60 T due to the Van Vleck effect approaching the \textit{J} = 3/2 state. The \textit{J} = 3/2 state, estimated to be 6 meV above \textit{J} = 1/2 from fits to inelastic neutron scattering data \cite{kim2021antiferromagnetic}, is evident in the Curie-Weiss fit to the magnetization between 150 and 300 K. From the $g$-factor anisotropy extracted from examining the magnetization saturation moments, we found a trigonal elongation that cannot be ignored in \NCTO{}. The resultant mixing between the two lowest doublet states leads to an easy-plane anisotropy, invalidating the pre-requisite for the Kitaev quantum spin liquid model. Additionally, we observed a distinct spin-flop-like phase transition leading to an unexpected 1/3 magnetization plateau-like phase, suggesting the existence of a weak Kitaev interactions. Therefore, we propose that the magnetic phase diagram can be elucidated by considering the XXZ model combined with Kitaev exchange interactions.

\section{Acknowledgement}
This work was principally lead by and supported by the U.S. Department of Energy, Office of Science, National Quantum Information Sciences Research Centers, Quantum Science Center. The facilities of the National High Magnetic Field Laboratory are supported by the National Science Foundation Cooperative Agreement No. DMR-2128556, and the State of Florida and the U.S. Department of Energy. Q. H. and H. D. Z. grew the samples with support from the National Science Foundation grant DMR-2003117. M. L. acknowledges the LDRD program at Los Alamos National Laboratory for initial work they did before joining the Quantum Science Center.



\renewcommand\thefigure{A\arabic{figure}}

\section{Appendix}
\setcounter{figure}{0}

In the appendix we present the full data set of magnetic susceptibility and magnetocaloric effect measurements with H $\parallel$ $c$-axis. In addition, we also present the magnetic susceptibility data with \textit{H} $\parallel a$-axis and the Curie-Weiss fitting parameters as a comparison to the c-axis parameters.

\begin{figure}[h]
\includegraphics[width=\linewidth]{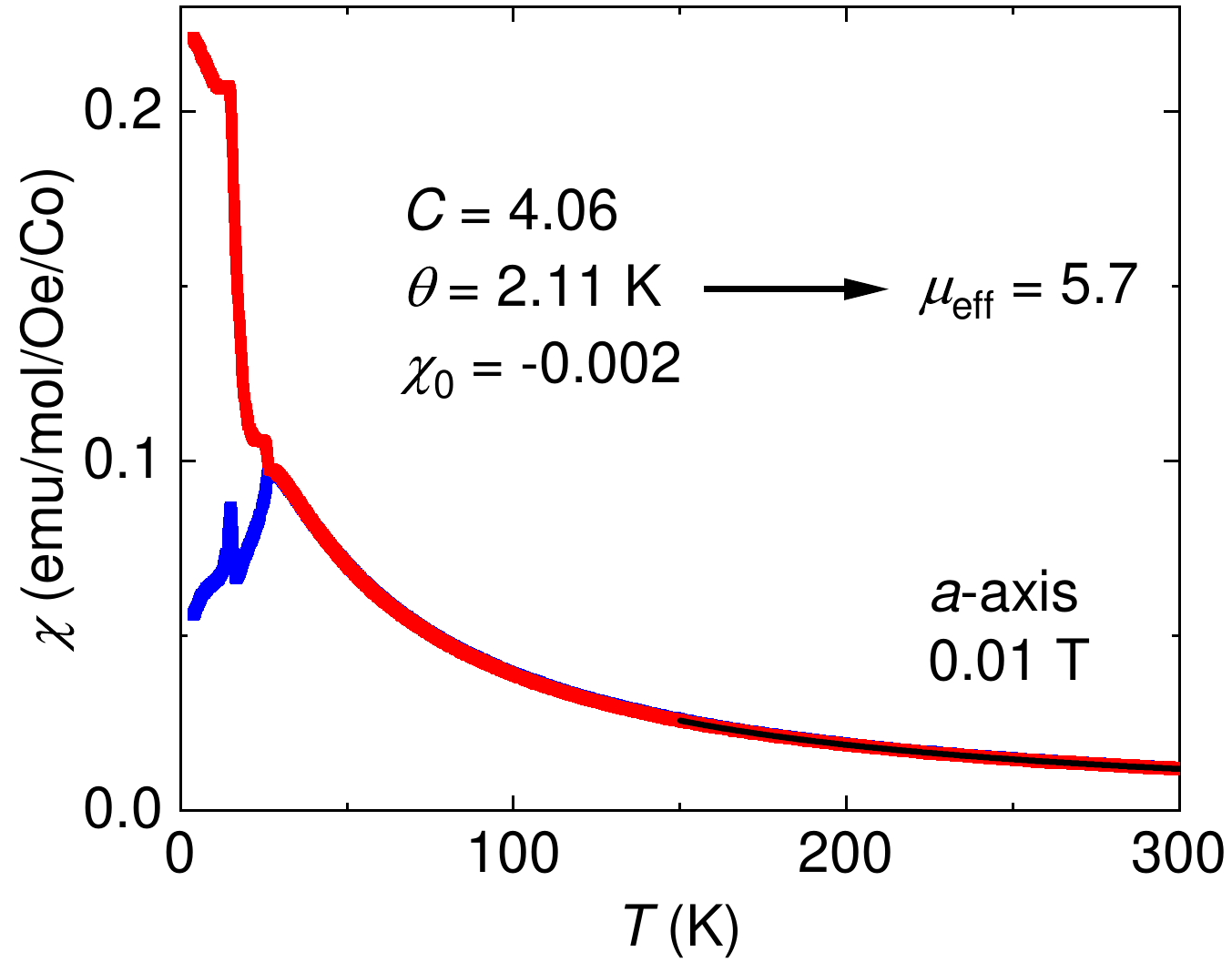}%
\caption{Magnetic susceptibility as a function of temperature at 0.01 T magnetic field. The thick red and blue curves are field-cooled and zero-field-cooled data, respectively. The thin black line is the Curie-Weiss fit from 150 K to 300 K. Extracted parameters and effective moments are shown in the panel. \label{Appendix_MvsT-a}}
\end{figure}

Fig.~\ref{Appendix_MvsT-a} shows the $a$-axis magnetic susceptibility as a function of temperature with the Curie-Weiss fitting parameters in the figure.

\begin{figure}[h]
\includegraphics[width=\linewidth]{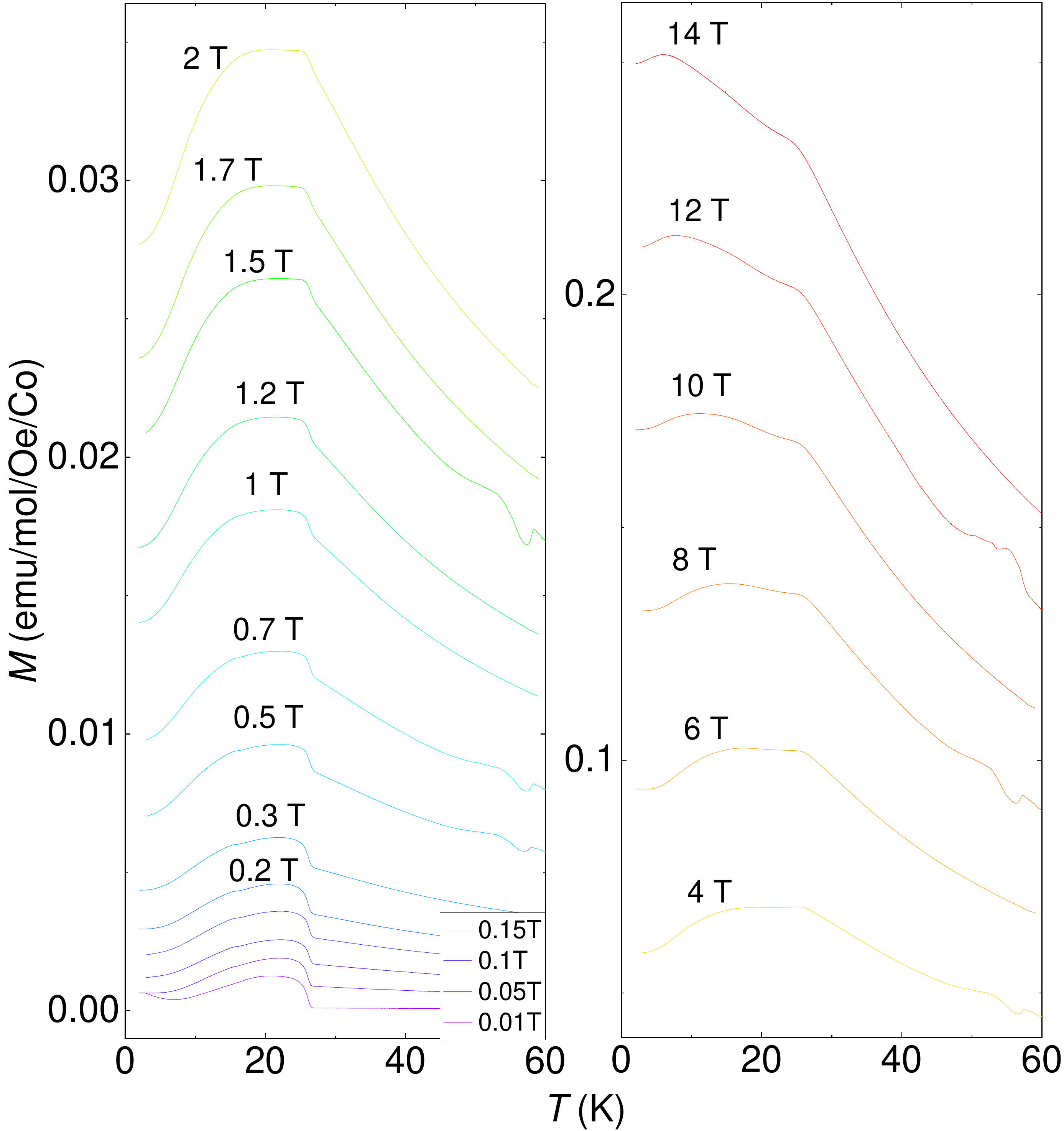}%
\caption{Full data set of magnetization as a function of temperature taken at various magnetic fields. The bottom four curves in left panel are taken at magnetic fields in the legend box. \label{Appendix_MvsT_full}}
\end{figure}

Fig.~\ref{Appendix_MvsT_full} shows the full data set of field-cooled magnetic susceptibility of \NCTO{} with \textit{H} $\parallel$ c-axis. 

\begin{figure}[h]
\includegraphics[width=\linewidth]{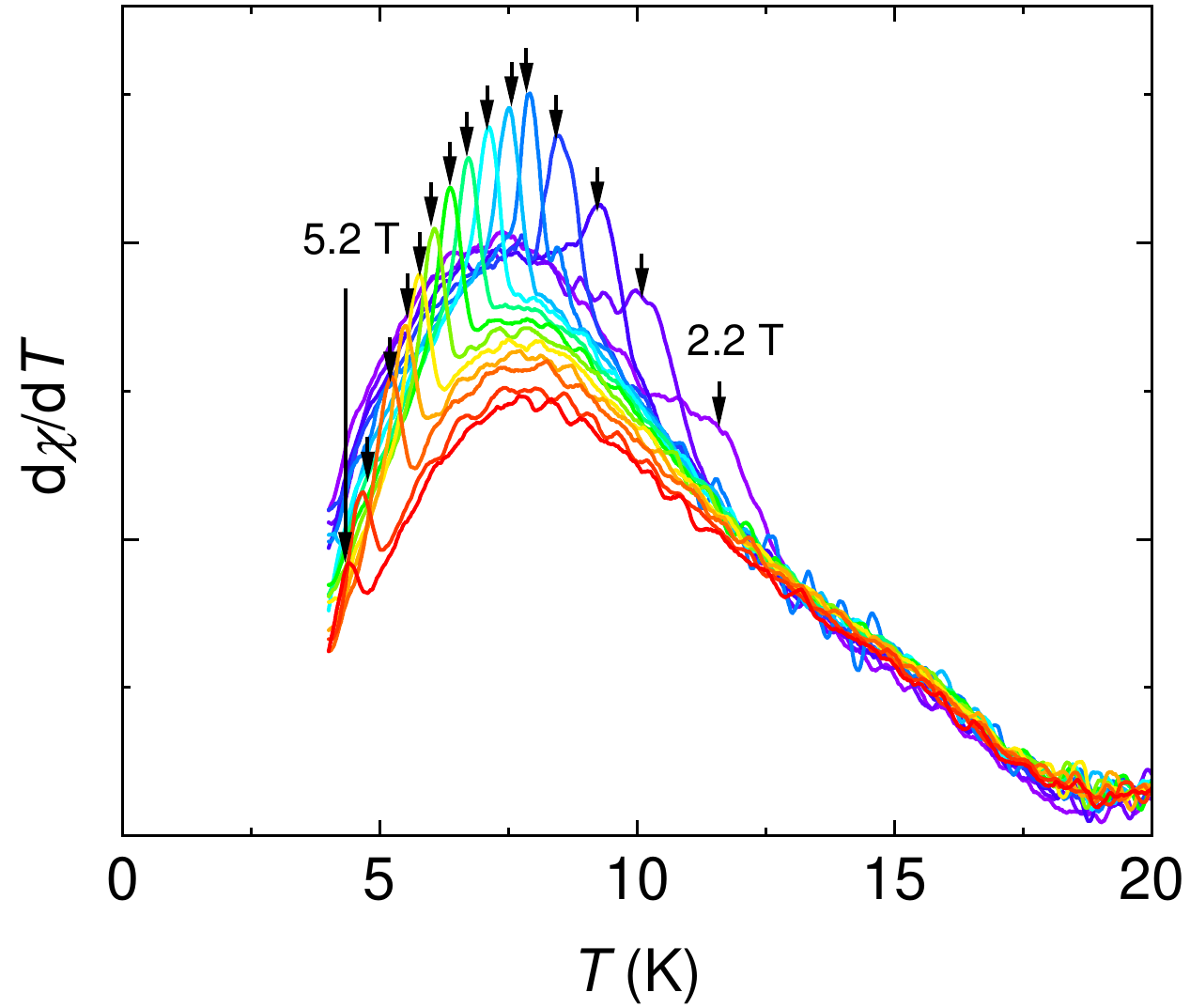}%
\caption{First derivative of magnetic susceptibility $d\chi/dT$ as a function of temperature. The transitions indicated by the black arrows are the ``open stars" recorded in the phase diagram of Fig.~\ref{PD} \label{Appendix-dchidT}}
\end{figure}

In Fig.~\ref{Appendix-dchidT} we highlight the feature corresponding to the open stars in Fig.~\ref{PD} by the black arrows. These are only observed in zero-field-cooled magnetic susceptibility vs. temperature measurements.

\begin{figure}[h]
\includegraphics[width=\linewidth]{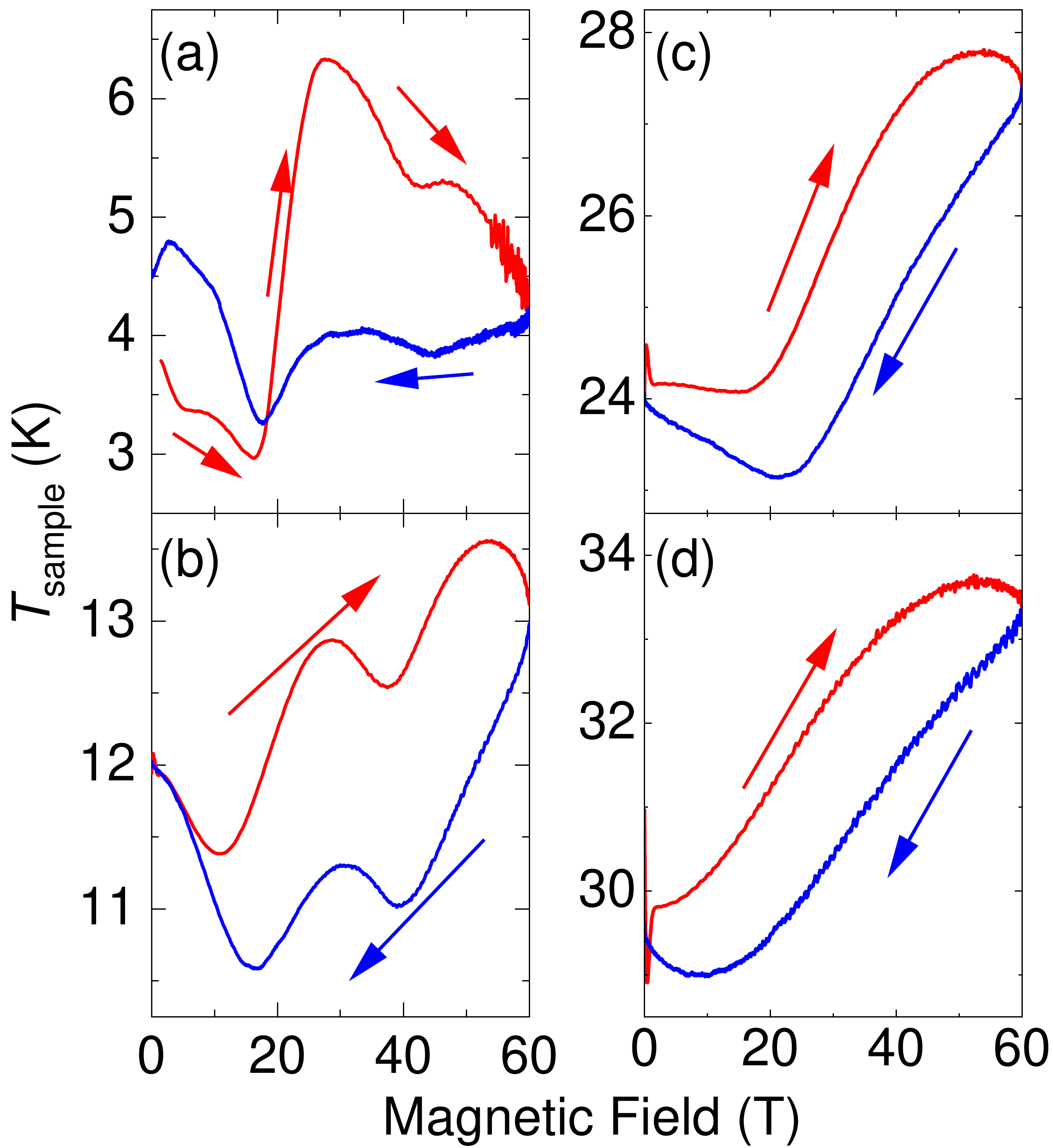}%
\caption{Sample temperatures as a function of magnetic field taken at (a) 3.92 K, (b) 12 K, (c) 24 K, and (d) 30 K. The red and blue curves are up- and down-field-sweeps, respectively, as indicated by the arrows. \label{Appendix-MC}}
\end{figure}

Fig.~\ref{Appendix-MC} demonstrates further the environmental cooling effect. Despite of all features preserved at certain magnetic fields at different temperatures ranging from 4 K to 30 K, in the field down-sweep they tend to be weaker, indicating the lost of adiabaticity. Notice also the data taken during field down-sweep are in general below the up-sweep data, remniscent of the environmental cooling effect. That is, when the field sweep rate is significantly lowered as we approached the peak field in up-sweep, the sample temperature started to relax towards the environmental temperature. 

\bibliography{NCTO}
\end{document}